\documentclass[%
    aps,
    prd,
    twocolumn,
    10pt,
    noeprint,
    amsmath,
    amssymb,
    superscriptaddress
]{revtex4-2}

\usepackage{graphicx}
\usepackage[%
    colorlinks=true,
    allcolors=blue
]{hyperref}
\usepackage[%
    print-unity-mantissa=false,
    range-phrase=-,
    range-units=single
]{siunitx}

\def\aap{Astron.\ Astrophys.}
\def\apjs{Astrophys.\ J.\ Suppl.\ Ser.}
\def\baas{Bull.\ Am.\ Astron.\ Soc.}
\def\cqg{Classical\ Quant.\ Grav.}
\def\cse{Comput.\ Sci.\ Eng.}
\def\epja{Eur.\ Phys.\ J.\ A}
\def\jors{J.\ Open\ Res.\ Software}
\def\joss{J.\ Open\ Source\ Software}
\def\mnras{Mon.\ Not.\ R.\ Astron.\ Soc.}
\def\nc{Nat.\ Commun.}
\def\nphys{Nucl.\ Phys.}
\def\prx{Phys.\ Rev.\ X}
\def\ptps{Prog.\ Theor.\ Phys.\ Suppl.}
\def\rmp{Rev.\ Mod.\ Phys.}

\begin{document}

\title{Problematic systematics in neutron-star merger simulations}

\author{F. Gittins}
\email{f.w.r.gittins@uu.nl}
\affiliation{Institute for Gravitational and Subatomic Physics (GRASP), Utrecht University, Princetonplein 1, 3584 CC Utrecht, Netherlands}
\affiliation{Nikhef, Science Park 105, 1098 XG Amsterdam, Netherlands}
\affiliation{Mathematical Sciences and STAG Research Centre, University of Southampton, Southampton SO17 1BJ, United Kingdom}

\author{R. Matur}
\affiliation{Mathematical Sciences and STAG Research Centre, University of Southampton, Southampton SO17 1BJ, United Kingdom}

\author{N. Andersson}
\affiliation{Mathematical Sciences and STAG Research Centre, University of Southampton, Southampton SO17 1BJ, United Kingdom}

\author{I. Hawke}
\affiliation{Mathematical Sciences and STAG Research Centre, University of Southampton, Southampton SO17 1BJ, United Kingdom}

\date{\today}

\begin{abstract}
    Next-generation gravitational-wave instruments are expected to constrain the equation of state of dense nuclear matter by observing binaries involving neutron stars. We highlight a problematic systematic error in finite-temperature merger simulations, where shock heating associated with the neutron-star surface gives rise to elevated temperatures. We demonstrate the severe implications of this artificial heating by computing static and dynamical tidal parameters for neutron stars immersed in simulation temperature profiles. The induced systematic errors must be addressed if we want to build robust gravitational-wave signal models for neutron-star, or indeed neutron star-black hole, binaries.
\end{abstract}

\maketitle

\section{Introduction}

Gravitational-wave astronomy has the potential to provide precise constraints on the extreme physics represented by neutron stars. The observation of the first binary neutron-star merger event, GW170817, in LIGO-Virgo highlighted this promise through the \textit{absence} of a distinguishable imprint of tidal interaction in the gravitational-wave signal \cite{2017PhRvL.119p1101A,2018PhRvL.121p1101A,2019PhRvX...9a1001A}. This, in turn, placed constraints on the neutron-star tidal deformability and the permitted range of neutron-star radii. To date, GW170817 is the only observed event for which such constraints have been obtained. This is expected to change with the development of more sensitive gravitational-wave interferometers, like the Einstein Telescope \cite{2010CQGra..27s4002P} and Cosmic Explorer \cite{2019BAAS...51g..35R}. Neutron-star physics provides a key science driver for these advanced instruments.

Designed to be about a factor of 50 more sensitive than the interferometers that were in operation in 2017, the next-generation detectors should allow us to put much tighter bounds on the equation of state of matter at supranuclear densities. This will involve precision measurements of the response of a neutron star to the tidal interaction with a companion during the late stages of compact binary inspiral. The tide induced on a neutron star has two aspects: At lower frequencies, the dominant feature is the \textit{static tide}, represented by the tidal deformability \cite{2008ApJ...677.1216H,2009PhRvD..80h4018B,2009PhRvD..80h4035D}. As the binary orbit shrinks and the system approaches merger, the \textit{dynamical tide} starts to play a role \cite{1994ApJ...426..688R,1994MNRAS.270..611L,2024PhRvD.109f4004P,2024PhRvD.109j4064H}. The dynamical tide has various features, but it is dominated by the contribution from the fundamental mode of oscillation of the star (the \textit{f}-mode) \cite{2020PhRvD.101h3001A}. Moreover, the tidal deformability and the \textit{f}-mode frequency are known to be linked by a \textit{quasi-universal relation} \cite{2014PhRvD..90l4023C}, which means that the two features may not be truly ``independent'' from the parameter-extraction point of view. There are convincing arguments that we need to include both static and dynamical aspects in our tidal models if we want to maximise the information extracted from observations \cite{2020NatCo..11.2553P,2022PhRvD.105l3032W,2022PhRvL.129h1102P}. 

Additional contributions to the dynamical tide---\textit{e.g.}, from low-frequency gravity \textit{g}-modes arising from composition and entropy gradients \cite{1994MNRAS.270..611L,2019MNRAS.489.4043A} or interface \textit{i}-modes associated with sharp phase transitions \cite{2012PhRvL.108a1102T,2021MNRAS.504.1129N}---are expected to leave a weak imprint on the gravitational-wave signal. Detections of such fine-print contributions would shed light on the composition and state of the high-density matter. This is an important issue, but not the focus of the discussion here.

Gravitational-wave astronomy relies, in general, on a robust link between observations and the underlying theory. In particular, for the problem of neutron-star tides, waveform models need to be calibrated against two distinct bodies of work \cite{2015PhRvL.114p1103B,2016PhRvL.116r1101H,2016PhRvD..94j4028S,2017PhRvD..96l1501D,2019PhRvD..99b4029D,2019PhRvD.100d4003D,2024PhRvD.109b4062A}. At low frequencies (early stages of inspiral), the problem is well described by perturbation theory coupled to a post-Newtonian model for orbital evolution. At higher frequencies---as the system approaches merger---the dynamics become non-linear and require numerical simulations. Each approach to the problem comes with its own set of computational issues. The specific feature we highlight here relates to the late stages of evolution where numerical simulations are necessary.

A variety of approaches are currently used to build waveform models for binaries involving neutron stars. For example, the time-domain effective-one-body framework \cite{2015PhRvL.114p1103B,2016PhRvL.116r1101H,2016PhRvD..94j4028S} calibrates the model using numerical-relativity simulations of merging black-hole binaries, while the frequency-domain closed-form tidal approximant method \cite{2017PhRvD..96l1501D,2019PhRvD..99b4029D,2019PhRvD.100d4003D,2024PhRvD.109b4062A}  normalises with respect to neutron-star simulations. There are also hybrid models that combine the two techniques, see for instance Ref.~\cite{2018PhRvD..97d4044K}. Arguably, the current state of the art relies on data from neutron-star merger simulations which implement fits of one-parameter nuclear-matter models \cite{2009PhRvD..79l4032R}, implicitly assuming the stellar material to be cold and the nuclear reactions to maintain equilibrium. However, realistic neutron-star mergers are hot, out-of-equilibrium events, so we ultimately need to work towards calibrations based on realistic, finite-temperature simulations \cite{2019EPJA...55..124P,2021PhRvD.104j3006H,2023ApJ...952L..36F} to capture the anticipated physics. Such simulations are essential if we want to connect the pre-merger behaviour with the violent dynamics, including electromagnetic counterpart emission, of the post-merger remnant.

\begin{figure}[ht]
    \includegraphics[width=\columnwidth]{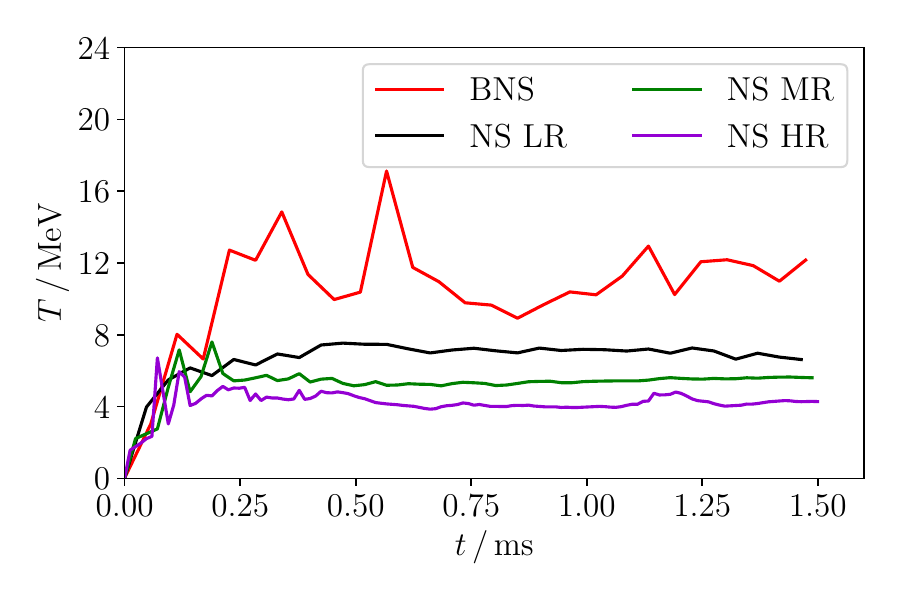}
    \caption{\label{fig:peakT}%
    The maximum temperature reached in the neutron-star matter during a numerical simulation across the entire domain. Results correspond to the DD2 equation of state and are provided for both single isolated neutron stars (at three different resolutions; \qty{369}{\metre} in black, \qty{185}{\metre} in green and \qty{92}{\metre} in purple) and one of the companions in a neutron-star binary simulation (with grid resolution \qty{221.5}{\metre} in red). In the binary simulation, the initial separation is set to \qty{45}{\kilo\metre}. The isolated neutron-star simulations suggest that the issue with the spurious heating remains for affordable levels of resolution. The comparison to the binary neutron-star case shows that the artificial heating is enhanced when the star is moving across the numerical grid. The temperatures indicated here are representative of those seen in all current finite-temperature (grid-based) neutron-star simulations.}
\end{figure}

The issue we bring to the fore relates to a (known, but somewhat ignored) systematic error present in all current (grid-based) finite-temperature neutron-star merger simulations involving shock-capturing schemes: The temperature in the simulated stars is much too high before merger.  The problem is due to the simple fact that the employed numerical schemes treat the sharp drop in density at the neutron-star surface as a ``shock'' leading to generation of (numerical) entropy and artificial heating. The level of heating varies between simulation codes, but it is generally the case that the temperature is ramped up to $\qtyrange{5}{10}{\mega\electronvolt}$ throughout the star very soon after the beginning of the simulation (see the evolution of the maximum temperature across the entire domain in Fig.~\ref{fig:peakT} and Appendix~\ref{app:Simulations}). As a result, the simulated neutron star is several orders of magnitude too hot. In fact, at this level the thermal pressure is significant, effectively making the star much less compressed than it ought to be. Based on a comparison to proto-neutron stars, which typically reach similar temperatures, one would estimate that the neutron-star radius can change by as much as a factor of 2. This obviously impacts on both the tidal deformability and the \textit{f}-mode frequency and hence may render debatable any waveform calibration involving such numerical simulation data. This is clearly problematic.

As a first step towards a better quantitative understanding of the issue, we calculate both static and dynamical tidal parameters for two specific realistic finite-temperature equations of state. Our results provide a more precise illustration of the impact of the artificial heating and a clear indication of the systematics associated with current finite-temperature neutron-star simulations.

\section{Systematics in finite-temperature simulations}

The fluid equations that determine the neutron-star structure are closed by a thermodynamical \textit{equation of state}. For our calculations, we assume a three-parameter model for the nuclear matter---for simplicity, only comprising neutrons, protons and electrons---with the temperature $T$, baryon-number density $n_\text{b}$ and electron fraction $Y_\text{e}$ as the natural thermodynamical variables. Such models represent the current state of the art in nuclear astrophysics \cite{2017RvMP...89a5007O} and are regularly implemented in numerical simulations \cite{2019EPJA...55..124P,2021PhRvD.104j3006H,2023ApJ...952L..36F}.

For the assumed three-parameter model, the first law of thermodynamics may be expressed as
\begin{equation}
    df = - n_\text{b} s \, dT + \frac{f + p}{n_\text{b}} \, dn_\text{b} + n_\text{b} \mu_\Delta \, dY_\text{e},
    \label{eq:first}
\end{equation}
where $s$ is the entropy per baryon, $p$ is the isotropic pressure, $\mu_\Delta = \mu_\text{p} + \mu_\text{e} - \mu_\text{n}$ and $\{ \mu_\text{n}, \mu_\text{p}, \mu_\text{e} \}$ are the neutron, proton and electron chemical potentials, respectively. The various quantities are as measured in the local inertial reference frame of the fluid. The free-energy density $f = f(T, n_\text{b}, Y_\text{e})$ represents the fundamental equation of state from which all the thermodynamical information about the stellar material can be derived.

In order to illustrate the artificial surface heating in binary simulations, we use temperature results from two separate inspiral-merger simulations with different prescriptions for the finite-temperature matter: APR \cite{2019PhRvC.100b5803S} and DD2 \cite{2010NuPhA.837..210H} (as implemented in the {\footnotesize CompOSE} library \cite{2015PPN....46..633T,2017RvMP...89a5007O,2022arXiv220303209T}). The first simulation is taken from Refs.~\cite{2021PhRvD.104j3006H,2021zndo...5469497H}, which uses the {\footnotesize Einstein Toolkit} codebase to collide neutron stars described by the APR nuclear-matter equation of state. The simulation clearly exhibits the characteristic artificial surface heating, reaching temperatures of order \qty{10}{\mega\electronvolt}, see Fig.~1 of Ref.~\cite{2021PhRvD.104j3006H}. In addition, mainly in order to demonstrate the behaviour with a different numerical set-up, we have performed simulations with the {\footnotesize WhiskyTHC} framework adopting the DD2 model for the microphysics. The details of the simulations are provided in Appendix~\ref{app:Simulations}. We consider two physical situations: an isolated neutron star and a binary merger for which the two stars (obviously) move across the numerical grid.

The results show that, at typical simulation resolutions, the bulk of an isolated neutron star heats up to temperatures of $\sim \qtyrange{1}{4}{\mega\electronvolt}$ (with the peak temperature slightly higher, see Fig.~\ref{fig:peakT}) due to fluid shocks on the grid. The evolution of the peak temperature reached in each simulation is displayed in Fig.~\ref{fig:peakT}. For the DD2 merger simulation, the results show that the artificial heating is further enhanced when the stars move across the numerical grid.

\section{Impact on tides}

The main question we want to answer is: What is the impact of the observed artificial heating on the main tidal parameters that one would aim to extract from gravitational-wave observations? In order to explore this issue, we solve for the linear perturbations of a spherically symmetric, perfect-fluid relativistic star. Our perturbation calculation closely follows Ref.~\cite{gittins2024neutronstar} and we provide further details in Appendix~\ref{app:Mode}. From the outset, we assume that the star is immersed in a temperature profile $T = T(n_\text{b})$. (This will be either uniform or lifted from a numerical simulation.) To determine the corresponding matter composition, we assume that the background fluid is in chemical equilibrium such that
\begin{equation}
    \mu_\Delta(T, n_\text{b}, Y_\text{e}) = 0.
\end{equation}
This fixes $Y_\text{e} = Y_\text{e}(n_\text{b})$ and reduces the thermodynamical state to depend solely on $n_\text{b}$. The structure of the equilibrium neutron star is then straightforwardly obtained by solving the standard equations supplemented with the functions $\varepsilon = \varepsilon(n_\text{b})$ and $p = p(n_\text{b})$ obtained from the equation of state. This provides the background on which the linear perturbations are computed. Although our set-up is similar to Ref.~\cite{gittins2024neutronstar}, here we are concerned with different temperature profiles, in particular those extracted from merger simulations.

As already mentioned, there are two regimes in a binary inspiral: the early regime, which is characterised by the static tide, and the late regime, where the dynamics become important (and non-linear aspects come into play). The static tide is commonly represented by the neutron star's \textit{tidal deformability} $\Lambda$. This quantity enters the inspiral waveform through its mass-weighted average for the binary at fifth order in the post-Newtonian approximation \cite{2008PhRvD..77b1502F,2017PhRvL.119p1101A}. The tidal deformability provides a dimensionless measure of the star's susceptibility to a companion's gravitational field. If the nuclear matter is particularly stiff, the neutron star can support large deformations (large values of $\Lambda$). Meanwhile,  if the stellar material is comparatively soft, the neutron star will be more compact and have a small $\Lambda$. These qualities make $\Lambda$ an attractive observable with which to constrain the properties of dense nuclear matter \cite{2008PhRvD..77b1502F}, as demonstrated in the case of GW170817 \cite{2017PhRvL.119p1101A,2018PhRvL.121p1101A,2019PhRvX...9a1001A}.

We determine the static, quadrupolar deformations using the formalism detailed in Ref.~\cite{2008ApJ...677.1216H} to obtain $\Lambda$. The required static perturbations depend on an additional aspect of the nuclear matter, the equilibrium adiabatic index
\begin{equation}
    \Gamma = \frac{\varepsilon + p}{p} \frac{dp}{d\varepsilon},
\end{equation}
which is determined from partial derivatives of the thermodynamical functions (see Ref.~\cite{gittins2024neutronstar}).

During the later, dynamical regime, the tidal driving frequency increases as the binary inspirals, radiating gravitational waves. As the orbital frequency increases, it will eventually become resonant with the low-frequency \textit{g}-modes of the neutron star, as well as implicate the other higher frequency oscillations, like the \textit{f}- and \textit{p}-modes \cite{1994ApJ...426..688R,1994MNRAS.270..611L}. Indeed, the \textit{f}-mode is expected to provide the dominant contribution to the tide \cite{2020PhRvD.101h3001A}. At this point, the assumption of a static tide breaks down and the problem becomes dynamical. In the absence of dissipation, the Newtonian oscillation problem is self-adjoint, implying that the modes form a complete basis in terms of which the tide can be decomposed. This  \textit{mode-sum} representation elegantly addresses the mathematical challenges of solving the full time-dependent, tidal response problem. Although motion in general relativity is inherently dissipative due to gravitational radiation, the natural vibrational modes of neutron stars are still expected to dominate the dynamical tide.

To represent the dynamical tide, we calculate the quadrupolar, quasi-normal oscillation modes of a neutron star using the equations provided in Refs.~\cite{1983ApJS...53...73L,1985ApJ...292...12D}. The equation of state now enters the fluid perturbations through the adiabatic index
\begin{equation}
    \Gamma_1 = \frac{\varepsilon + p}{p} \left( \frac{\partial p}{\partial \varepsilon} \right)_{s, Y_\text{e}},
\end{equation}
assuming frozen composition during an oscillation (\textit{i.e.}, that the mode dynamics are fast enough that we may ignore nuclear reactions). Similarly to $\Gamma$, the determination of $\Gamma_1$ involves thermodynamical derivatives (for more detail, see Ref.~\cite{gittins2024neutronstar}).

The two adiabatic indices $\Gamma$ and $\Gamma_1$ characterise stratification in the fluid. In general, stellar material can support both entropy and composition gradients. When $\Gamma_1 \geq \Gamma$, the star is convectively stable and supports low-frequency \textit{g}-mode oscillations. Meanwhile, when $\Gamma_1 = \Gamma$, the \textit{g}-modes vanish. For realistic nuclear-matter models, the neutron star is expected to be stably stratified throughout, except possibly at low densities, and will therefore support  \textit{g}-mode oscillations. Indeed, a recent study has shown that these low-frequency fluid oscillations, which likely become resonant with the tide as the binary inspirals, may be within reach of next-generation observatories \cite{2023PhRvD.108d3003H}.

We present a sample of numerical results in Table~\ref{tab:Tide}. For each of the two nuclear-matter models we consider, we show the tidal parameters (static and dynamical) when the star has cold, uniform temperature and when the star is placed in a temperature profile obtained from a numerical-relativity merger simulation (modelled on the results in Fig.~\ref{fig:profile} in Appendix~\ref{app:Simulations} for DD2), assuming the same matter model (and stars with the same baryon mass). This provides an immediate quantitative measure of the impact of the artificial simulation temperatures on the tidal parameters.

\begin{table}[ht]
    \caption{\label{tab:Tide}%
    The effects on the tidal parameters due to artificial temperatures in numerical-relativity merger simulations. The table lists the stellar radius $R$, tidal deformability $\Lambda$, \textit{f}-mode frequency $\omega_\textit{f}$ and first three \textit{g}-mode frequencies $\omega_\textit{g$_1$}$, $\omega_\textit{g$_2$}$ and $\omega_\textit{g$_3$}$. For each nuclear-matter model, APR and DD2, the parameters are shown for two neutron stars with different temperatures; one neutron star has uniform temperature, whereas the other is immersed in a merger simulation profile. The stars described by APR have baryon mass $M_\text{b} = 1.40 M_\odot$ and those for DD2 have $M_\text{b} = 1.55 M_\odot$. It is evident that the neutron stars with simulation-level temperatures have significantly altered tidal parameters compared to the cold stars.}
    \sisetup{table-format=4.2}
    \vspace*{0.25cm}
\begin{tabular}{ l S S S S }
    \hline\hline
    & \multicolumn{2}{c}{APR} & \multicolumn{2}{c}{DD2} \\
    \cline{2-5}
    & \multicolumn{1}{c}{\qty{0.02}{\mega\electronvolt}} & \multicolumn{1}{c}{Simulation} & \multicolumn{1}{c}{\qty{0.2}{\mega\electronvolt}} & \multicolumn{1}{c}{Simulation} \\
    \hline
    $R$/\unit{km} & 11.6 & 25.1 & 13.2 & 22.9 \\
    $\Lambda$ & 459.4 & 534.4 & 680.9 & 852.8 \\
    $\text{Re}[\omega_\textit{f} / (2 \pi)]$/\unit{\hertz} & 1912.3 & 1598.9 & 1607.7 & 1526.6 \\
    $\text{Re}[\omega_\textit{g$_1$} / (2 \pi)]$/\unit{\hertz} & 508.5 & 856.4 & 277.6 & 622.3 \\
    $\text{Re}[\omega_\textit{g$_2$} / (2 \pi)]$/\unit{\hertz} & 304.5 & 672.8 & 138.0 & 533.7 \\
    $\text{Re}[\omega_\textit{g$_3$} / (2 \pi)]$/\unit{\hertz} & 126.1 & 579.4 & 114.9 & 416.4 \\
    \hline\hline
\end{tabular}
\end{table}

For the APR matter model, the numerical results show that $\Lambda$ is 16\% larger due to the artificially high temperature in the numerical simulation than for the corresponding cold neutron star. The difference is even starker with DD2, for which the deformability increases by more than 25\% with the simulation temperature profile. It should be noted that the two equation-of-state cases are not intended to be directly compared to each other. Indeed, the involved neutron stars have different masses, as well as different numerical treatments for the hydrodynamics and the spacetime evolution. Rather, the point of the analysis is to illustrate that the artificial temperatures in two \textit{separate} merger simulations lead to the same qualitative behaviour: The tidal parameters are significantly distorted.

The results for the tidal deformability are quite intuitive. The artificially high simulation temperatures contribute to a strong thermal pressure, which leads to larger stellar radii. In fact, from Table~\ref{tab:Tide}, we see that the neutron-star radii approximately double. This leads to an effective stiffening of the nuclear matter and means that $\Lambda$---which scales with the areal radius $R$ as $\propto R^5$---increases. It is not particularly surprising that higher temperatures lead to larger tidal deformabilities (this feature is generic). What is notable is the extent to which the values change. If we were to calibrate gravitational-waveform models with these kinds of simulations, these aspects would inevitably manifest as systematic errors that bias parameter inference.

We now turn our attention to the effect on the dynamics. Presented in Table~\ref{tab:Tide} are the (real) oscillation frequencies of the \textit{f}-mode and first three \textit{g}-modes. The results show that these are also drastically altered by the artificial temperature. We see that, when the neutron star is placed in the temperature profile of a simulation, the spectrum shifts considerably. The \textit{f}-mode is the most weakly affected of the oscillations, decreasing by 16\% for the APR model, but only 5\% for DD2. In contrast, the low-frequency \textit{g}-modes are substantially altered, increasing in frequency due to enhanced entropy gradients caused by the high temperatures.

\begin{figure*}[ht]
    \includegraphics[width=0.5\textwidth]{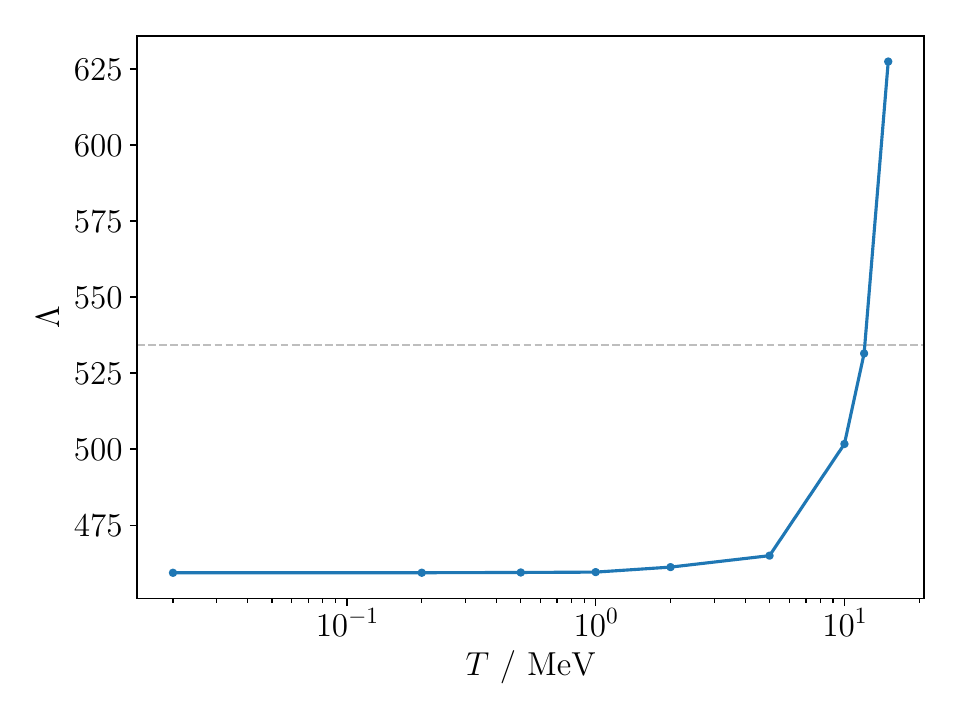}%
    \includegraphics[width=0.5\textwidth]{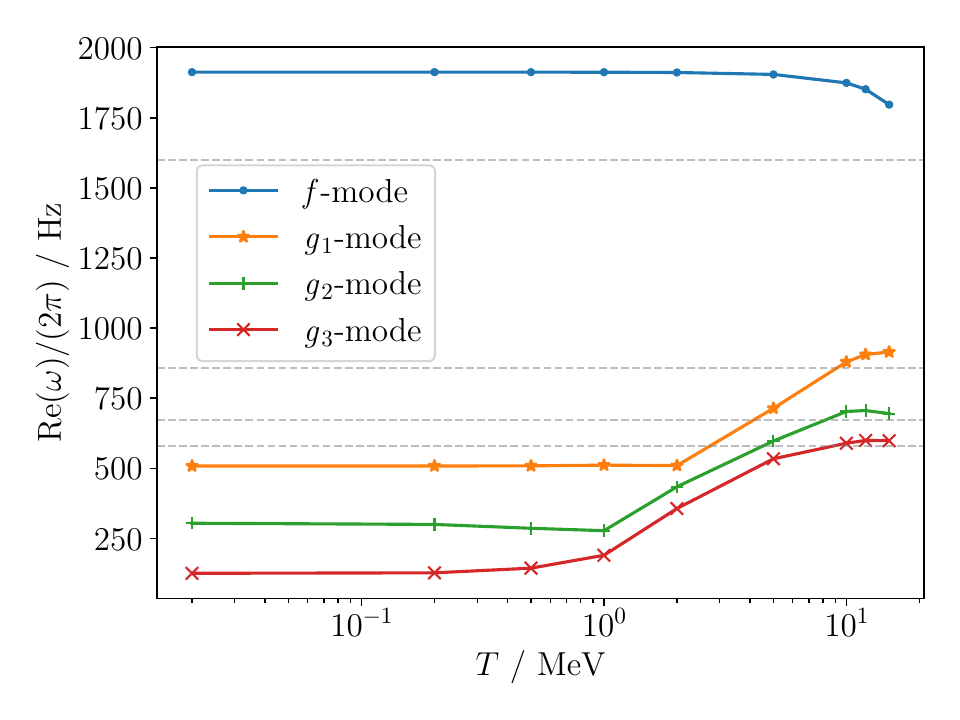}
    \caption{\label{fig:Summary}%
    The tidal deformability (left panel) and mode eigenfrequencies (right panel) against uniform temperature of an $M_\text{b} = 1.40 M_\odot$ neutron star described by the APR equation of state. The frequencies correspond to the \textit{f}- and first three \textit{g}-modes, as indicated by the legend. The horizontal, dashed lines correspond to the parameters listed in Table~\ref{tab:Tide} calculated using the simulation temperature profile. As the neutron star is heated up, the \textit{f}-mode frequency decreases with larger neutron-star radius. The other quantities, including the tidal deformability and \textit{g}-mode frequencies, increase.}
\end{figure*}

Finally, we calculate the tidal parameters' evolution with respect to uniform temperature. The results are presented in Fig.~\ref{fig:Summary} for APR nuclear matter and are consistent with the behaviour we have just discussed: The tidal deformability and \textit{g}-mode frequencies increase with temperature, while the \textit{f}-mode gradually oscillates slower. We see that the \textit{f}-mode is the least affected by temperature. In each panel of Fig.~\ref{fig:Summary}, we show the tidal values from Table~\ref{tab:Tide} using the simulation profile for comparison. The tidal deformability and \textit{g}-mode frequencies in the simulation are very roughly similar to that of a star with uniform $T \sim \qty{10}{\mega\electronvolt}$. However, even at $T = \qty{15}{\mega\electronvolt}$, the frequency of the \textit{f}-mode remains higher than that of the simulated neutron star.

\section{Words of caution}

We have demonstrated that the key tidal parameters of neutron stars in finite-temperature, merger simulations are severely distorted due to artificial heating arising from numerical shocks close to the stellar surface. Using data from two numerical-relativity simulations, with different numerical methodologies and matter prescriptions, we showed how quantities associated with both the static and dynamical tide are substantially shifted by the artificially high temperatures. Specifically, this includes the neutron-star tidal deformability, already constrained by the GW170817 observation, as well as the oscillation modes that dominate the dynamical tide, which we hope to probe with upcoming, next-generation gravitational-wave interferometers, the Einstein Telescope and Cosmic Explorer. Although we considered only two simulations, we expect these features to be generic for all grid-based numerical-relativity codes, as they all reach unphysically high temperatures. The alternative strategy to quench the thermal evolution during the inspiral phase and reactivate it close to merger~\cite{2016PhRvD..94d4060K}, may to some extent reduce the problem, but it introduces less controlled errors when the temperature is switched on.

The main take-away message from this work is cautionary. If we were to use results from finite-temperature non-linear simulations to calibrate gravitational waveforms, we may incur considerable systematic error in the parameter inference. Current instruments are not sensitive enough to the tide for these issues to manifest, but the problem will need to be addressed if we want to make precision observations with the Einstein Telescope and Cosmic Explorer.

Future work will need to be dedicated to either reducing the temperature systematics or correcting for them in the gravitational-wave analyses. As a first step, we need to better understand the origin of the problem. This includes reconciling the results we have presented here with previous work using simpler equation-of-state models (like polytropes) and a phenomenological thermal ($\Gamma$-law) representation. For example, the results in Ref.~\cite{2000MNRAS.313..678F} demonstrate much better agreement between simulations and perturbation theory. To some extent, the difference in the results likely stems from the fact that the phenomenological equation-of-state models assume chemical equilibrium. To illustrate this, consider Eq.~\eqref{eq:first} as a relation for the numerical errors. By enforcing chemical equilibrium we have $\mu_\Delta = 0$ and the final term does not contribute. Away from chemical equilibrium we would need the error $dY_\text{e}$ to ``match'' the other errors to give a sensible temperature. However, in current simulations the evolved quantity is (roughly) $n_\text{b} Y_\text{e}$, and hence the accuracy will be poor at low densities. This is likely related to the observation---see for example the results presented in Fig.~3 of \cite{2021PhRvD.104j3006H}---that finite-temperature equation-of-state simulations are prone to large errors at low temperatures. Of course, these arguments only hint at the origin of the problem we are discussing. We do not yet have a solution.

In order to make progress, it would be helpful to understand to what extent the issue we have raised is relevant for particle-based simulations, like those in Refs.~\cite{2014PhRvD..90b3002B,2024arXiv240415952R}. In principle, such simulations provide a better representation of the neutron-star surface, but it remains to be established to what extent this alleviates the artificial heating issue.

Let us make two final remarks. First, and this is important in order to keep the discussion in the proper context, the problem we have discussed does \textit{not} influence current gravitational-waveform models for neutron-star tides. These models, like the work in Ref.~\cite{2024PhRvD.109b4062A}, do not (yet) involve thermodynamically consistent matter models and hence do not suffer the artificial heating problem (at least not at the level we have indicated here). Having said that, if we want to do better in the future (and we do), then we need to get a handle on the temperature problem. Second, the artificial heating of low-density matter may have considerable impact on the related problem of matter ejecta, for which finite-temperature simulations \textit{are} being used \cite{2022CQGra..39a5008N}. In this problem, the additional thermal pressure may tend to unbind matter and an elevated temperature will affect the composition (the electron fraction) of the outflows, which may in turn alter the nuclear reaction rates and the associated kilonova signature. Similarly, an artificially high temperature will impact on neutrino opacities, and hence need to be carefully considered in efforts to implement realistic neutrino transport \cite{2024PhRvL.132u1001E}. Whether this concern is justified remains to be explored.

\begin{acknowledgments}
    FG acknowledges funding from the European Union (ERC, DynTideEOS, 101151301). NA and IH acknowledge support from STFC via grant number ST/R00045X/1. The authors thank Peter Hammond for sharing temperature data from a numerical-relativity merger simulation and Tim Dietrich and Nick Stergioulas for useful discussions. We acknowledge the use of the IRIDIS High Performance Computing Facility and associated support services at the University of Southampton. The software developed to support this article is available in a GitHub repository \cite{code} and is written in the {\footnotesize Julia} programming language \cite{Bezanson:2014pyv,DifferentialEquations.jl-2017,Interpolations.jl,2024arXiv240316341P,Optim.jl-2018}. The figures were generated using {\footnotesize Matplotlib} \cite{2007CSE.....9...90H,PythonCall.jl}.
    
    Views and opinions expressed are however those of the authors only and do not necessarily reflect those of the European Union or the European Research Council. Neither the European Union nor the granting authority can be held responsible for them.
\end{acknowledgments}

\appendix

\section{\label{app:Simulations}Numerical simulations}

For the finite-temperature inspiral-merger simulations we considered two distinct models. The first simulation used the APR matter model \cite{2021PhRvD.104j3006H,2021zndo...5469497H} and was performed using the {\footnotesize Einstein Toolkit} \cite{2012CQGra..29k5001L}. The initial data was created using {\footnotesize Lorene} \cite{lorene}, while the hydrodynamical and spacetime evolution was performed using {\footnotesize GRHydro} \cite{2005PhRvD..71j4006H,2005PhRvD..71b4035B,2007CQGra..24S.235G,2014CQGra..31a5005M} and {\footnotesize McLachlan} \cite{mclachlan,2009PhRvD..79d4023B}, respectively. {\footnotesize McLachlan} uses the BSSN formulation \cite{1987PThPS..90....1N,1995PhRvD..52.5428S,1998PhRvD..59b4007B} of the Einstein equations. The simulation was performed by setting the atmosphere temperature and rest-mass density to \qty{0.02}{\mega\eV} and \qty{6.2e6}{\gram\per\centi\metre\cubed}, respectively. A fourth-order Runge-Kutta method was employed, with the Courant-Friedrichs-Lewy condition set to $0.25$. Neutron stars were tracked using {\footnotesize NSTracker}. (For more information on the implementation see Ref.~\cite{2021PhRvD.104j3006H}.)

The second set of simulations, for the DD2 matter model, involved initial data generated using the {\footnotesize FUKA} solver \cite{2021PhRvD.104b4057P} and the evolution of the hydrodynamics and spacetime were performed with {\footnotesize WhiskyTHC} \cite{2012A&A...547A..26R,2014MNRAS.437L..46R,2014CQGra..31g5012R,2015ASPC..498..121R} and {\footnotesize CTGamma} \cite{2011PhRvD..83d4045P}, respectively. We used the {\footnotesize Z4C} formulation \citep{2010PhRvD..81h4003B} of the Einstein equations, the {\footnotesize Carpet} adaptive mesh refinement driver (as in the APR model) \cite{2004CQGra..21.1465S} of {\footnotesize Cactus} \cite{805282}, and tracked the neutron stars with the {\footnotesize BNSTrackerGen} component of {\footnotesize WhiskyTHC}. While we performed the hydrodynamical evolution using the finite volume method, the Einstein equations were evolved using a 4th-order finite difference method. We used the local Lax-Friedrichs flux splitting method, and for the reconstruction, we applied a 5th-order monotonicity-preserving scheme. The time evolution was carried out using a fourth-order Runge-Kutta method and the Courant-Friedrichs-Lewy condition was set to $0.15$. The atmosphere temperature and rest-mass density were set to \qty{0.02}{\mega\eV} and \qty{6.2e3}{\gram\per\centi\metre\cubed}, respectively. We used the Sophie Kowalevski release of the {\footnotesize Einstein Toolkit} \cite{2012CQGra..29k5001L,2022zndo...7245853H}.

To explore the effects of heating, we considered both single and binary neutron-star simulations. In these simulations, we set the individual ADM masses of the neutron stars to $1.44 M_\odot$ and considered an equal-mass system for the binary case. All simulations that used the DD2 equation of state in this study employed a cell-centred grid structure.

For the binary neutron-star simulation, we extended the domain to $(x, y, z) = (\qty{2835}{\kilo\metre}, \qty{2835}{\kilo\metre}, \qty{1418}{\kilo\metre})$ (applying reflection symmetry along the $z$-axis). We used $8$ refinement levels and set the finest grid to \qty{221}{\metre}. The radius of the finest refinement level in both regions was \qty{15}{\kilo\metre}, covering the neutron stars, and we re-gridded the innermost refinement levels every 128 iterations to track the motion of the neutron stars. In this case, the gravitational-wave frequency was \qty{581}{\hertz} initially and \qty{726}{\hertz} at \qty{3.9}{\milli\second} after the beginning of the simulation (approximately $1$ orbit out of $3.5$), when the temperature profile was created. We used the \texttt{$1+$log} and {\footnotesize Gamma Driver} gauge conditions and set the constraint damping coefficients to $\kappa_{1} = 0.02$ and $\kappa_{2} = 0$ for the binary neutron-star merger.

For the single-star models, we used the same numerical methods and only modified the grid setup. We performed three simulations using different resolutions. In these cases, we extended the domain to $(x, y, z) = (\qty{89}{\kilo\metre}, \qty{89}{\kilo\metre}, \qty{89}{\kilo\metre})$ (applying reflection symmetry along all axes) and used adaptive mesh refinement with $4$ refinement levels. The finest grid resolutions were set to \qty{369}{\metre}, \qty{185}{\metre}, and \qty{92}{\metre}. The radius of the finest refinement level was \qty{15}{\kilo\metre}, covering the neutron stars in each case.

\begin{figure*}[ht]
    \includegraphics[width=\textwidth]{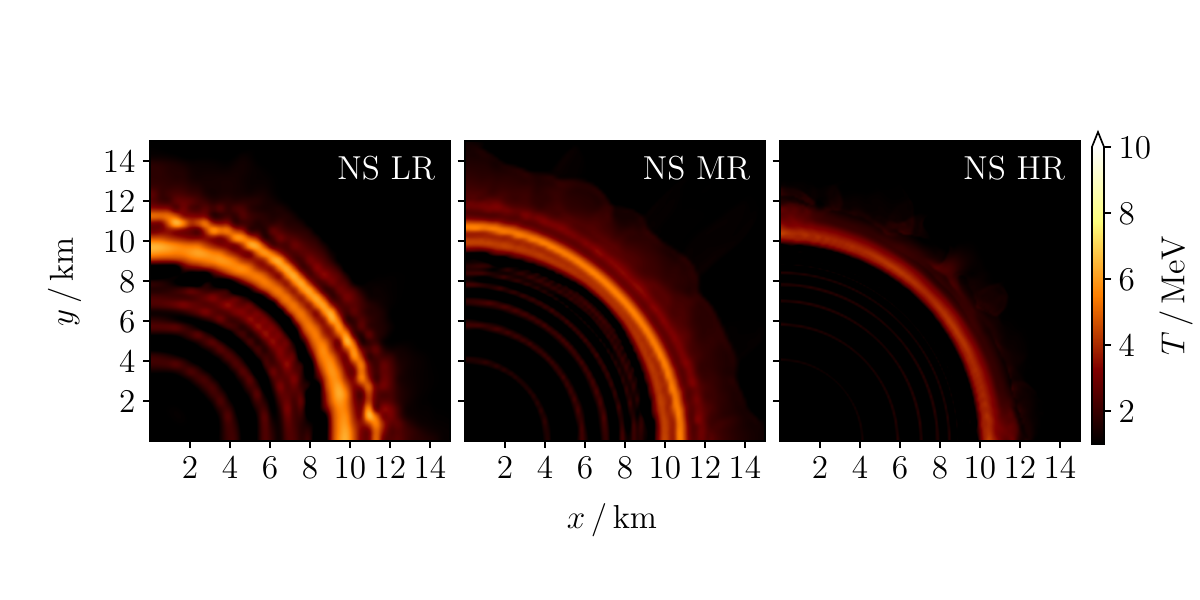}
    \caption{\label{fig:shock}%
    An illustration of the artificial heating associated with shocks at the neutron-star surface. The results are obtained from the evolution of a single isolated neutron star and the three panels represent different simulation resolutions. The left, middle and right panels show the low-resolution (LR), medium-resolution (MR) and high-resolution (HR) simulation results, with a grid spacing of $\sim \qty{369}{\metre}$,  $\sim \qty{185}{\metre}$ and $\sim \qty{92}{\metre}$ in the finest grid, respectively. The neutron stars are modelled using the finite-temperature, composition-dependent DD2 equation of state. While the temperature in the core of the  NS reaches $\sim \qty{4}{\mega\electronvolt}$ in the low-resolution simulation, it reduces to $\sim \qtyrange{1}{2}{\mega\electronvolt}$ for the high-resolution simulation within nearly $\sim \qty{1.6}{\milli\second}$.}
\end{figure*}

In Fig.~\ref{fig:shock}, we present the artificial heating for simulations of isolated (static) neutron stars and three different numerical resolutions, corresponding to scales of \qty{369}{\metre}, \qty{185}{\metre} and \qty{92}{\metre}, respectively. The results show that the star heats up to temperatures of $\sim \qtyrange{1}{4}{\mega\electronvolt}$ due to fluid shocks on the grid. It is notable that, while the shocks originate at the surface, the heat rapidly propagates to the stellar interior as the simulation progresses.

\begin{figure}[ht]
    \includegraphics[width=\columnwidth]{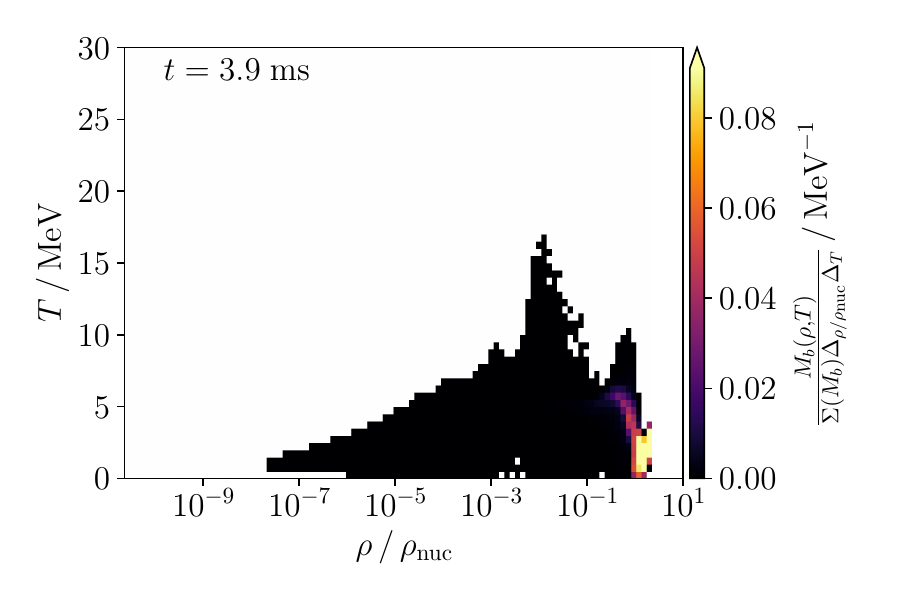}
    \caption{\label{fig:profile}%
     The typical temperature profile (in the equatorial plane) extracted from the DD2 binary neutron-star simulation. The histogram illustrates the distribution of baryon mass in the density-temperature plane. The data is extracted after \qty{3.9}{\milli\second} of evolution.}
\end{figure}

The inspiral-merger simulation shows that the artificial heating is enhanced when the stars move across the numerical grid.  The typical temperature profile extracted from our DD2 simulation, demonstrating that the peak temperatures reaches well above \qty{10}{\mega\electronvolt}, is shown in Fig.~\ref{fig:profile}.

To create the temperature profile (Fig.~\ref{fig:profile}) for the binary neutron-star simulation, we used three-dimensional data for the rest-mass density, temperature, velocity and metric components. Subsequently, we computed the baryonic mass by mapping this data onto a fixed grid and generated a two-dimensional histogram of the rest-mass density and temperature, weighted by the baryonic mass, to analyse their distributions.

\section{\label{app:Mode}Perturbative mode calculations}

The mode calculations closely follow the analysis in Ref.~\cite{gittins2024neutronstar}. Specifically, to address noise with low-frequency oscillations, we use an augmented approach developed in Ref.~\cite{2015PhRvD..92f3009K} and perturbations in the exterior are calculated using the method from Ref.~\cite{1995MNRAS.274.1039A}. A mode with discrete (complex) frequency $\omega$ corresponds to a solution such that the ingoing gravitational waves in the exterior, with amplitude $\tilde{A}_\text{in}$, vanish.

\begin{figure*}[ht]
    \includegraphics[width=\columnwidth]{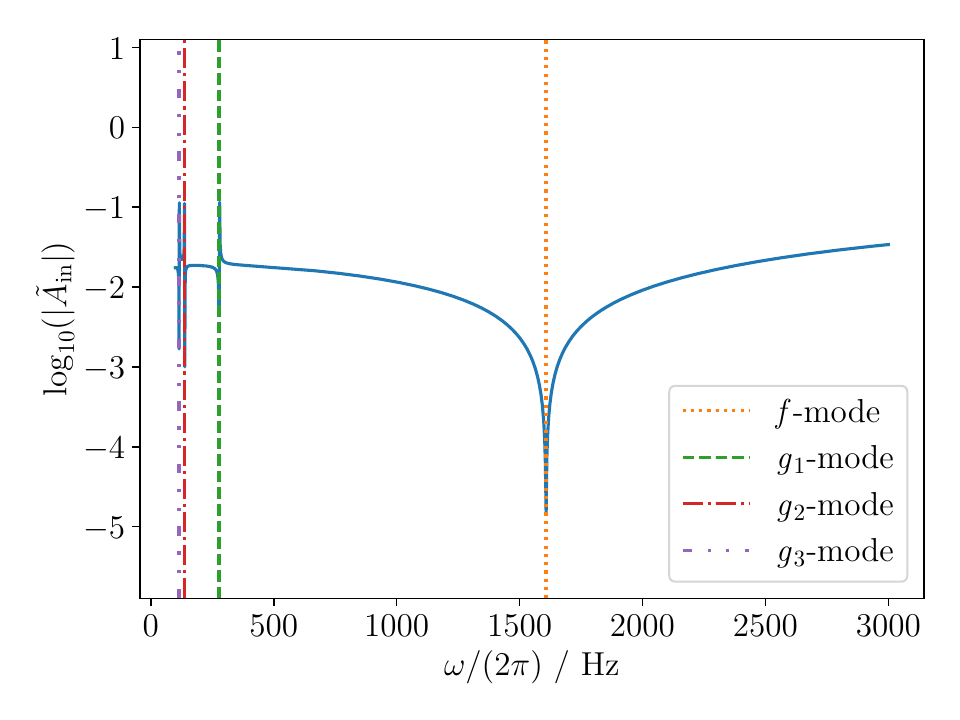}%
    \includegraphics[width=\columnwidth]{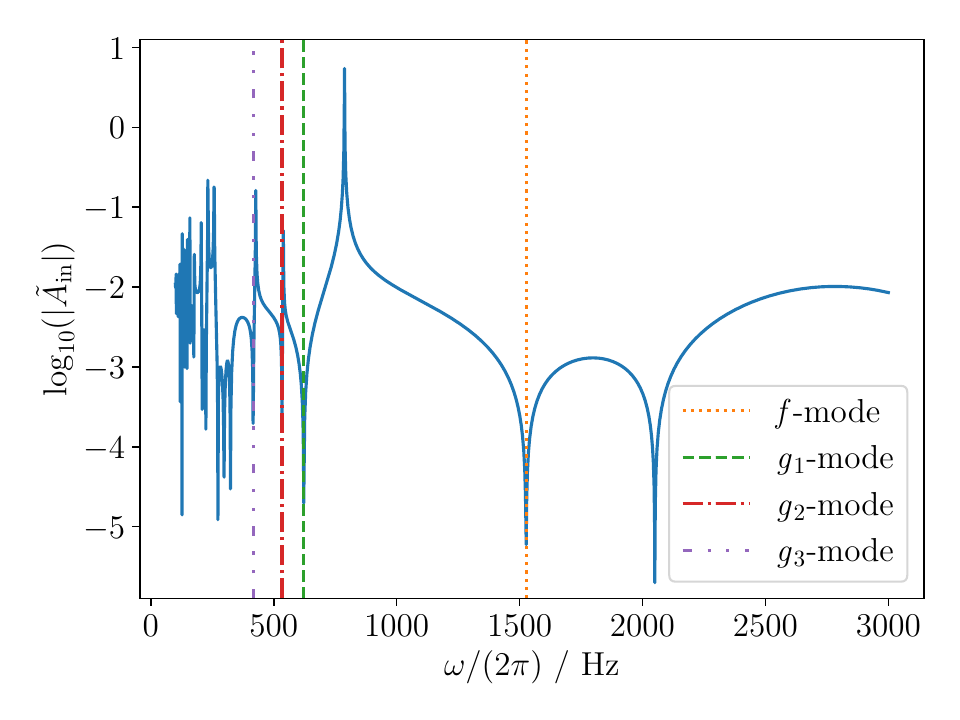}
    \caption{\label{fig:DD2}%
    The oscillation spectra of an $M_\text{b} = 1.55 M_\odot$ neutron star described by the finite-temperature DD2 equation of state for two different temperature profiles; uniform temperature $T = \qty{0.2}{\mega\electronvolt}$ (left panel) and the simulation profile (right panel). Here, $\tilde{A}_\text{in}$ represents the amplitude of ingoing gravitational radiation and $\omega$ is the (real) frequency of the perturbations. A mode of the neutron star corresponds to when $\tilde{A}_\text{in} = 0$. The weakly damped oscillations can be identified by the singularities in the spectrum. (Rapidly damped, spacetime \textit{w}-modes are not visible.) The vertical lines indicate the mode eigenfrequencies listed in Table~\ref{tab:Tide}. The high temperatures in the merger simulation lead to a substantial distortion of the mode spectrum. The \textit{f}- and \textit{p}-mode frequencies decrease. We see in the right panel how the first \textit{p}-mode becomes visible, with frequency $\text{Re}[\omega / (2 \pi)] = \qty{2050.2}{\hertz}$. The impact of the simulation temperatures on the \textit{g}-modes is to increase their frequencies.}
\end{figure*}

The mode-calculations show that, at low temperatures in the APR neutron star, the first \textit{g}-mode---the \textit{g$_1$}-mode---is an interfacial \textit{i}-mode that arises due to the core-crust phase transition \cite{gittins2024neutronstar}. Such oscillations appear when there is a discontinuity in the energy density of the equation of state. This identification is made clear by examining the mode's eigenfunctions, which possess a sharp cusp at the point in the star where the discontinuity lies. It is notable that this mode oscillates at a relatively high frequency of $\text{Re}[\omega/(2 \pi)] = \qty{508.5}{\hertz}$. To illustrate the changes, we show in Fig.~\ref{fig:DD2} the neutron-star oscillation spectra for the finite-temperature DD2 equation of state. The qualitative results for APR, summarised in Table~\ref{tab:Tide}, are the same.

\bibliography{refs.bib}

\end{document}